\documentclass[runningheads]{llncs}
\usepackage{textgreek}
\usepackage{csquotes}
\usepackage{multirow}
\usepackage{booktabs}
\usepackage[tableposition=top]{caption}
\usepackage{enumitem}
\usepackage{amsmath}
\usepackage{longtable}
\usepackage{perpage} 
\MakePerPage{footnote} 
\usepackage{ulem}
\usepackage{graphicx}
\usepackage[colorlinks=true, allcolors=blue]{hyperref}
\usepackage{subfiles}

\usepackage{xcolor}

\begin{document}
\title{Explaining Search Result Stances to Opinionated People}

%
%

\author{Zhangyi Wu\inst{1}\orcidID{0009-0001-2540-9967} \and
Tim Draws\inst{2}\orcidID{0000-0001-5053-4674} \and
Federico Cau\inst{1}\orcidID{0000-0002-8261-3200} \and
Francesco Barile\inst{1}\orcidID{0000-0003-4083-8222} \and Alisa Rieger\inst{2}\orcidID{0000-0002-2274-1606} \and
Nava Tintarev\inst{1}\orcidID{0000-0003-1663-1627} 
}
\authorrunning{Wu et al.}
%
\institute{Maastricht University, Maastricht, Netherlands \and
Delft University of Technology, Delft, Netherlands\\
\email{\{zhangyi.wu\}@student.maastrichtuniversity.nl,\\
\{federico.cau,f.barile,n.tintarev\}@maastrichtuniversity.nl,\\
\{t.a.draws,a.rieger\}@tudelft.nl}
}
%
\maketitle              
\begin{abstract}
People use web search engines to find information before forming opinions, which can lead to practical decisions with different levels of impact. The cognitive effort of search can leave opinionated users vulnerable to cognitive biases, e.g., the \textit{confirmation bias}. In this paper, we investigate whether stance labels and their explanations can help users consume more diverse search results. We automatically classify and label search results on three topics (i.e., \textit{intellectual property rights}, \textit{school uniforms}, and \textit{atheism}) as \textit{against}, \textit{neutral}, and \textit{in favor}, and generate explanations for these labels. In a user study ($N$=203), we then investigate whether search result stance bias (balanced vs biased) and the level of explanation (plain text, label only, label and explanation) influence the diversity of search results clicked. We find that stance labels and explanations lead to a more diverse search result consumption. However, we do not find evidence for systematic opinion change among users in this context.
We believe these results can help designers of search engines to make more informed design decisions.

\end{abstract}
\section{Introduction}

Web search that can lead to consequential decision-making frequently concerns \textit{debated topics}, topics that different people and groups disagree on, such as \textit{whether to vaccinate a child} or \textit{whether nuclear energy should be used as a power source}. 
Prior research has shown that the interplay between search engine biases and users' cognitive biases can lead to noteworthy behavioral patterns. 
For instance, when search result stance biases interact with cognitive user biases, information seekers may experience the \textit{search engine manipulation effect} (SEME): the tendency to adopt the stance expressed by the majority of (highly-ranked) search results~\cite{allam2014ImpactSearchEngine,bink2022FeaturedSnippetsTheir,epstein2015SearchEngineManipulation,pogacar2017PositiveNegativeInfluence}. However, these results have only been studied and found for users who are undecided, not for users who already have strong opinions, who we refer to as \textit{opinionated} users. 

High cognitive demand during complex web searches can increase the risk of cognitive biases~\cite{azzopardi2021}. One such bias is the \textit{confirmation bias}, which involves a preference for information that aligns with preexisting beliefs while disregarding contradictory information during the search process~\cite{azzopardi2021,nickerson1998}.
Interventions to mitigate confirmation bias during web search have aimed at decreasing engagement with attitude-confirming and increasing engagement with attitude-opposing information~\cite{rieger2021ThisItemMight}, i.e., reducing interaction with search results that confirm a user's attitude and increasing interaction with search results that challenge a user's attitude.
Interventions to reduce interaction with particular items have also been investigated in the context of misinformation.
One effective method for reducing interaction with misleading content involves \textit{labels} to flag certain items~\cite{chen2023,kaiser2021,mena2020}.

The core issue with confirmation bias during web search, similar to the related issues of misinformation and SEME, is that users consume biased content. This motivated us to investigate interventions to increase the diversity of consumption and specifically whether labels indicating stance, and their explanations, are likewise successful for confirmation bias mitigation during search on debated topics. 
Therefore the goal of these interventions is to promote unbiased web search and mitigate the effects of users' confirmation bias and underlying (stance) biases in a \textit{search engine result page} (SERP).
Consequently, this paper aims to address the following question: \textit{\textbf{Do automatically generated stance labels and explanations of the labels for search results increase the diversity of viewpoints users engage with, even if search results are biased?}}

To address this question, we test three hypotheses. Previous work has found that undecided users are likely to change their opinion when exposed to biased search results, since they select more search results reflecting a certain opinion~\cite{allam2014ImpactSearchEngine,draws2021ThisNotWhat,epstein_search_2015,pogacar2017PositiveNegativeInfluence}.
However, in this study we restrict participants to opinionated users, having \textit{strong} existing opinions on a topic, and investigate whether \textit{\textbf{H1a):} Users who are exposed to viewpoint-biased search results interact with \underline{less diverse results} than users who are exposed to 
 balanced search results.}

Second, informative labels have been shown to mitigate confirmation bias in search results~\cite{rieger2021ThisItemMight}. Therefore, in this study, we investigate whether simple stance labels (\textit{against}, \textit{neutral}, and \textit{in favor}), and stance labels with explanations (importance of keywords) are effective for mitigating confirmation bias. This leads to \textit{\textbf{H1b)}: Users who are exposed to search results with (1) stance labels or (2) stance labels with explanations for each search result interact with more diverse content than users who are exposed to regular search results}.

Third, if the labels are effective in reducing confirmation bias, we would expect an interaction effect between the bias in search results and explanation level (plain, label, label plus explanation): \textit{ \textbf{H1c)} Users who are exposed to search results with (1) stance labels or (2) stance labels with explanations are less susceptible to the effect of viewpoint biases in search results on clicking diversity.}

We investigate these hypotheses in a pre-registered between-subjects user study ($N$=203) simulating an open search task.\footnote{The pre-registration is openly available at \url{https://osf.io/3nxak}.}
Our results show that both stance labels and explanations, led to a more diverse search result consumption compared to plain (unlabeled) search result pages. However, we did not find evidence that the explanations influenced opinion change. We believe these results can help designers of search engines to make more informed design decisions.

\section{Related Work}

Explainable Artificial Intelligence (XAI) aims to help people understand the decisions and predictions AI systems make. In this paper, we investigate specifically how XAI can support users in searching for disputed topics. Search for debated topics is highly subjective: when users search the web to seek advice or form opinions on these kinds of topics, not just search result \textit{relevance} but also the stance of content is influential~\cite{allam2014ImpactSearchEngine,draws2021ThisNotWhat,epstein2015SearchEngineManipulation,pogacar2017PositiveNegativeInfluence,rieger2021ThisItemMight}. To mitigate undesired effects such as biased opinion change, earlier work has measured and increased the fairness~\cite{gezici2021EvaluationMetricsMeasuring,yang2017MeasuringFairnessRanked,zehlike2021FairnessRankingSurvey} and viewpoint diversity in search results~\cite{draws2023ViewpointDiversitySearch,puschmann2019BubbleAssessingDiversity,white2013BeliefsBiasesWeb}.

On the user interface side, it could be fruitful to label and explain the stance represented on a search engine results page (or SERP). These labels are related to the task known as \textit{stance detection}, which is predominantly applied in a \textit{target-specific} fashion. That is, detecting not just a sentiment, but how it is referred to in relation to a specific topic or claim (often referred to as the \textit{target}, e.g., ``people should wear \underline{school uniforms}'')~\cite{aldayel2021StanceDetectionSocial}. Stance detection is a multi-class classification task (i.e., typically classifying documents into \textit{against}, \textit{neutral}, and \textit{in favor}, so predictive performances are most commonly reported in terms of macro F1  scores~\cite{kucuk2021StanceDetectionSurvey}. Furthermore, web search interventions targeting the mitigation of undesired effects, such as SEME, require \textit{cross-target} stance detection models to quickly respond to the large variety of debated topics users may search for. Here, stance detection models are applied to data sets where each document may be relevant to one of many potential topics~\cite{aldayel2021StanceDetectionSocial,kucuk2021StanceDetectionSurvey}. Constructing models that classify documents into stances related to \textit{any} topic in such a way may lead to weaker predictive accuracy compared to target-specific methods, but makes stance detection more generalizable and scalable. Cross-target ternary stance detection by previous work (e.g., on news articles or tweets) have ranged roughly from macro F1 scores of $.450$ to $.750$~\cite{aldayel2019your,allaway-mckeown-2020-zero,augenstein2016StanceDetectionBidirectional,hardalov2022few,reuver2021StanceDetectionTopicIndependent,xu-etal-2018-cross}. 
 
Also comparable are the cross-topic stance detection models evaluated using the \textit{Emergent} data set (and its follow-up version, the \textit{2017 Fake News Challenge} data set) which have achieved macro F1 scores of up to $.756$~\cite{hanselowski-etal-2018-retrospective,roy2022ExploitingStanceHierarchies,sepulveda2021exploring}.
While the main contribution of this paper is not to improve on the state of the art for stance detection, the stance detection method (DistilBERT) used here is comparable to this state of the art (macro F1 of 0.72). DistilBERT is much smaller than other pre-trained models, and handles small datasets well \cite{silalahi2022named,staliunaite2020compositional,tong2022multimodel}.

What XAI methods are suitable for explaining stance detection to users? Stance detection can be seen as a text classification task. For text classification, explanations containing \textit{input features} have been found to be highly adaptable and often meaningful to humans~\cite{draws2023explainable,madsen2022post}.

The way in which explanations can be visualized depends on the data type, purpose, and audience. Many current methods indicate input features as feature importance using a saliency map~\cite{Ribeiro2016WhySI,IG_Sundararajan2017AxiomaticAF}.
When the features are readable texts, saliency information is shown by highlighting the most significant input at word or token level \cite{jin2019bridging}. 
There have been some instances where researchers used text-based saliency maps to demonstrate their findings \cite{gohel2021explainable}.
To the best of our knowledge, no previous work has explored whether highlighting salient words in search results would mitigate people's clicking bias.
One of the most closely related works found that feature-based explanations could help users simulate model predictions for search results \cite{draws2023explainable}.
Another similar study involves a verbal saliency map using a model-agnostic explainer and a human evaluation of explanation representations of news topic classifier and sentiment analysis \cite{feldhus2022constructing}. Their finding is that the saliency map makes explanations more understandable and less cognitively challenging for humans than heatmap visualization.
However, our work differs from theirs in several ways: we study explanations in the context of search engines, and we have conducted a full-fledged user study while they only performed a pilot study.

\textbf{Contribution to Knowledge for XAI.}
Previous XAI literature has contributed to explaining information retrieval systems, focusing on the interpretability of document-retrieval mechanisms \cite{text_rank2023,Lyu2023list,Yu2022leg}. 
For example, the authors of \cite{Yu2022leg} propose a listwise explanation generator, which provides an explanation that covers all the documents contained in the page (e.g., by describing which query aspects were covered by each document). These explanations were not evaluated by people. 
Another paper studied how well explanations of individual search results helped people anticipate model predictions \cite{draws2023explainable}, but did not consider cognitive or stance bias.
In contrast to previous work, this paper examines how explanations affect users, considering the potential mitigation of their cognitive biases on search engine manipulation effects (SEME). 
 In doing so, we see a need to address both potential (stance) bias within a search result page, and the bias of users consuming these results. 
 To the best of our knowledge, this is also the first work to conduct empirical experiments on users' \textit{behavior} in response to explanations in the context of information retrieval.

\section{Methodology}
This section describes the materials we used for organizing the user study (e.g., data set, stance detection model, explanation generation, and search interface).

\subsection{Data Preparation}
\label{sec:datapreparation}

\begin{table}[ht]
    \centering
    \setlength{\tabcolsep}{12pt}
    \renewcommand{\arraystretch}{1.3}
    \begin{tabular}{llc}
         \toprule
        & & \textbf{Stance Distribution} \\
        \textbf{Topic} & \textbf{N} & Against -- Neutral -- In Favor \\
         \hline
        \small{Intellectual property rights} & 378 & 10.5\% -- 17.7\% -- 71.7\% \\
        \small{School uniforms} & 395 & 21.5\% -- 36.7\% -- 41.8\% \\
        \small{Atheism} & 352 & 19.8\% -- 46.3\% -- 33.8\% \\
        \hline
        \textbf{Total} & 1125 & 17.3\% -- 33.3\% -- 49.3\% \\
         \bottomrule
    \end{tabular}
    \caption{Topic and stance distribution in the used data.}
    \label{tab:data}
\end{table}

To train, test, and explain the stance detection model, we considered a public data set containing search results related to three debated topics (i.e., \textit{atheism}, \textit{intellectual property rights}, and \textit{school uniforms})~\cite{draws2023viewpointdiversity}.\footnote{The data set is available at \url{https://osf.io/yghr2}.} 
\cite{draws2023viewpointdiversity} motivate the selection of these three topics because they offer valid arguments for both supporting and opposing viewpoints. Additionally, they argue that opinions on these topics have diverse impacts, ranging from concerning mainly the user (atheism) to businesses (intellectual property rights) and society (school uniforms).
These data include URLs, titles, snippets, and stance labels for a total of 1475 search results, which had been retrieved via API or web crawling from two popular web search engines.

Stance labels had been assigned (by experts) on seven-point Likert scales (i.e., including three degrees of opposing or supporting a topic), which we mapped into the three categories \textit{against}, \textit{neutral}, and \textit{in favor} (i.e., which is what most current stance detection methods handle).
Using the provided URLs, we crawled the full web page text bodies (stripped of any HTML tags) for all search results.
We here dropped 347 search results from the data as their text bodies could not be retrieved (e.g., because of 404 errors), leaving 1125 search results accompanied by their respective text bodies.
Finally, we processed the retrieved contents by truncating each document's middle section, retaining only its head and tail, then concatenating each search result's title, snippet, and the head section and tail section, while ensuring that the result is exactly 510 tokens long.
We removed all other information from the data aside from the documents' stance labels.
Table \ref{tab:data} shows the stance distribution per topic in our final data set.

\subsection{Stance Detection Model}
\label{sec:model}

\begin{figure}
    \centering
    \includegraphics[width=0.78\textwidth]{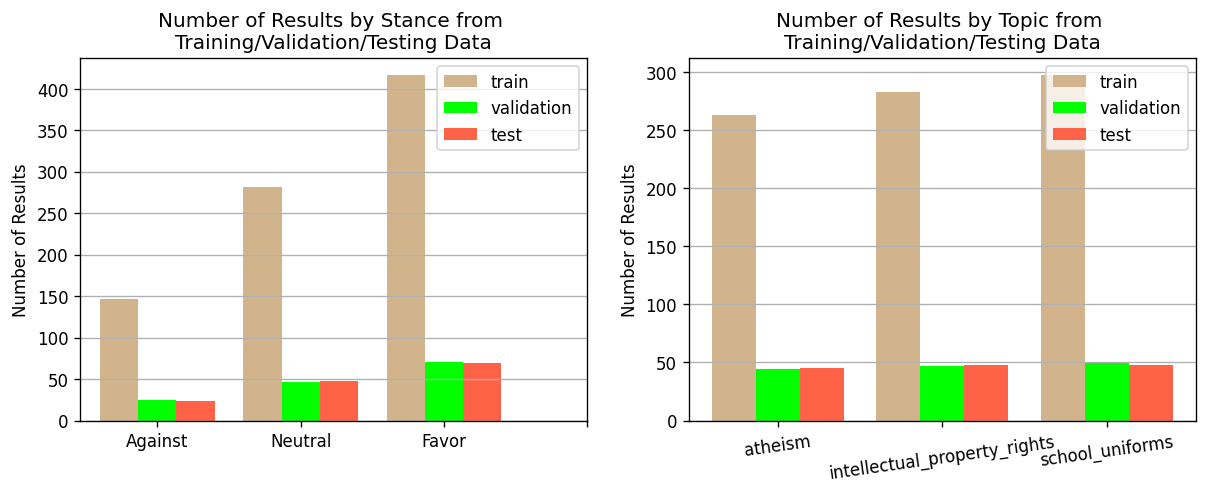}
\caption{Distributions of Attributes Across the Training, Validation and Test Set}
    \label{fig:distribt}
\end{figure}

After pre-processing (tokenization), we developed the model for classifying search results into against, neutral, and in favor. The dataset was split into training (75\%) and validation/test (25\%) sets. 
We fine-tuned the model for different hyperparameters. 
After every epoch the validation set were evaluated to monitor its learning progress. 
Once the model's evaluation loss stops to decrease for 5 epochs the training is terminated and  is evaluated based on the unseen test set, and the result is considered the general performance that we report. 
The best-performing model and learned parameters were used as the predictor for identifying labels and generating explanations.

\subsubsection{Tokenization.}

Before training the stance classification model, we needed further preprocessing of the raw search results. Specifically, we had to \textit{tokenize} each word before feeding the search results into the model. 
The tokenization process performs tasks such as handling subwords, adding special tokens to pad a sequence of words to a \texttt{max length}, or indicating the beginning or end of a sequence.
In our work, we instantiated the DistilBERT tokenizer using the AutoTokenizer class provided by the \texttt{transformers} \cite{wolf-etal-2020-transformers}.
We set the \texttt{length} of the tokenizer to a size of 512.

\subsubsection{Training Details.}
Similar to previous work performing search result stance classification~\cite{draws2023explainable}, we built one cross-topic model using our entire data set. First, we split the dataset in a stratified manner, which preserves the percentage of the topic and labels in each subset. Figure \ref{fig:distribt} shows the final split across the training, validation, and test sets on both topics and labels.
Then, we classified search results into stance categories (i.e., \textit{in favor}, \textit{against}, or \textit{neutral}), using a pre-trained version of the uncased DistilBERT base model \cite{Sanh2019DistilBERTAD} from HuggingFace. Specifically, we fine-tuned DistilBERT using 75\% (843) of the documents in our data and split the remaining 25\% (282) equally for validation and testing. 
Due to the relatively small size of our dataset, we tried to avoid over-fitting by using neural network dropout and stopping training when evaluation loss stops improving for 5 epochs~\cite{srivastava2014dropout,ying2019overview}.
We trained using the same dataset split and experimented with learning rates ranging from $5e\textendash6$ to $1e\textendash4$.
Regarding the remain hyper-parameters, the models were optimized using the Adam optimizer with a learning rate of $1e-5$, the batch size was set to 8 and we set the dropout rate of both attention layers and fully connected layers to 0.4.

\subsubsection{Metric.}
In our stance detection task, we have three labels to be classified, and their distribution is uneven.
To take performance on all stances into account, we considered the macro F1 score, defined as the average of the F1 scores for each of the three stances:
\begin{equation}
\small
macro_{F1} = ({F1}(stance=favor) + {F1}(stance=neutral) + {F1}(stance=against))/3
\end{equation}

\subsubsection{Model performance.}
We obtained a stance detection model by fine-tuning the DistilBERT model for our downstream classification task. The fine-tuning process was completed through HuggingFace's \texttt{Trainer} interface. The progress of each run was tracked using Weights \& Biases.\footnote{\url{https://wandb.ai}} We observed that the learning rate of $1e\textendash5$ gave the best performance with a macro F1 score of 0.72.\footnote{Due to a minor error in evaluation, a slightly higher macro F1 score was reported in the pre-registration. However, this erroneous score did not influence the training process or affect our user study.}

\subsection{Creating Explanations}
\label{sec:lime}
\subsubsection{Instance Selection.}
For the user study, we selected search results correctly predicted by the model, picking them mainly from the test set and some from the validation set.
As mentioned before, the train, validation, and test sets were split in a stratified way, which preserves the frequencies of our target classes. To assemble search results for our study, we randomly drew 21 \textit{correctly predicted} search results per topic from our test and validation data (i.e., seven against, seven neutral toward, and seven in favor of the topic). The SERPs later displayed 10 results per page. 

\subsubsection{LIME Parameter Tuning.}
After we fine-tuned and evaluated the model on the search result corpus, we used LIME (Local Interpretable Model-Agnostic Explanations) to explain the model's predictions \cite{Ribeiro2016WhySI}.\footnote{\url{https://github.com/marcotcr/lime}} 
For text data, LIME outputs a list of features (tokens) from the input sentence, ranked by their importance w.r.t. a model's specific prediction. We generated the explanations by setting the neighborhood size to learn the linear model to 5000, kernel\_width to 50,\footnote{We tried multiple kernel sizes (10, 25, 50, and 75) and chose a value of 50 since we got a slight increase in the R2 scores for each LIME local prediction on the test set of about 3-4\% on average compared to the other sizes.} and showing the top 20 important tokens (or less based on the text length) belonging to the model's predicted class. 

\subsection{Search Engine}
\label{sec:sepp}

\subsubsection{Architecture.}
We implemented our web-based search interface using the SearchX platform as a basis \cite{putra2018searchx}. The web application has both a front-end and a back-end.
The front-end is built on NodeJS using React and Flux frameworks and manages users' data by sending logs to the back-end, which mainly handles the retrieval of search results from the database and stores the logs from the front-end.  

\subsubsection{Interface.}

\begin{figure}[!ht]
    \centering
    \includegraphics[width=0.6\textwidth]{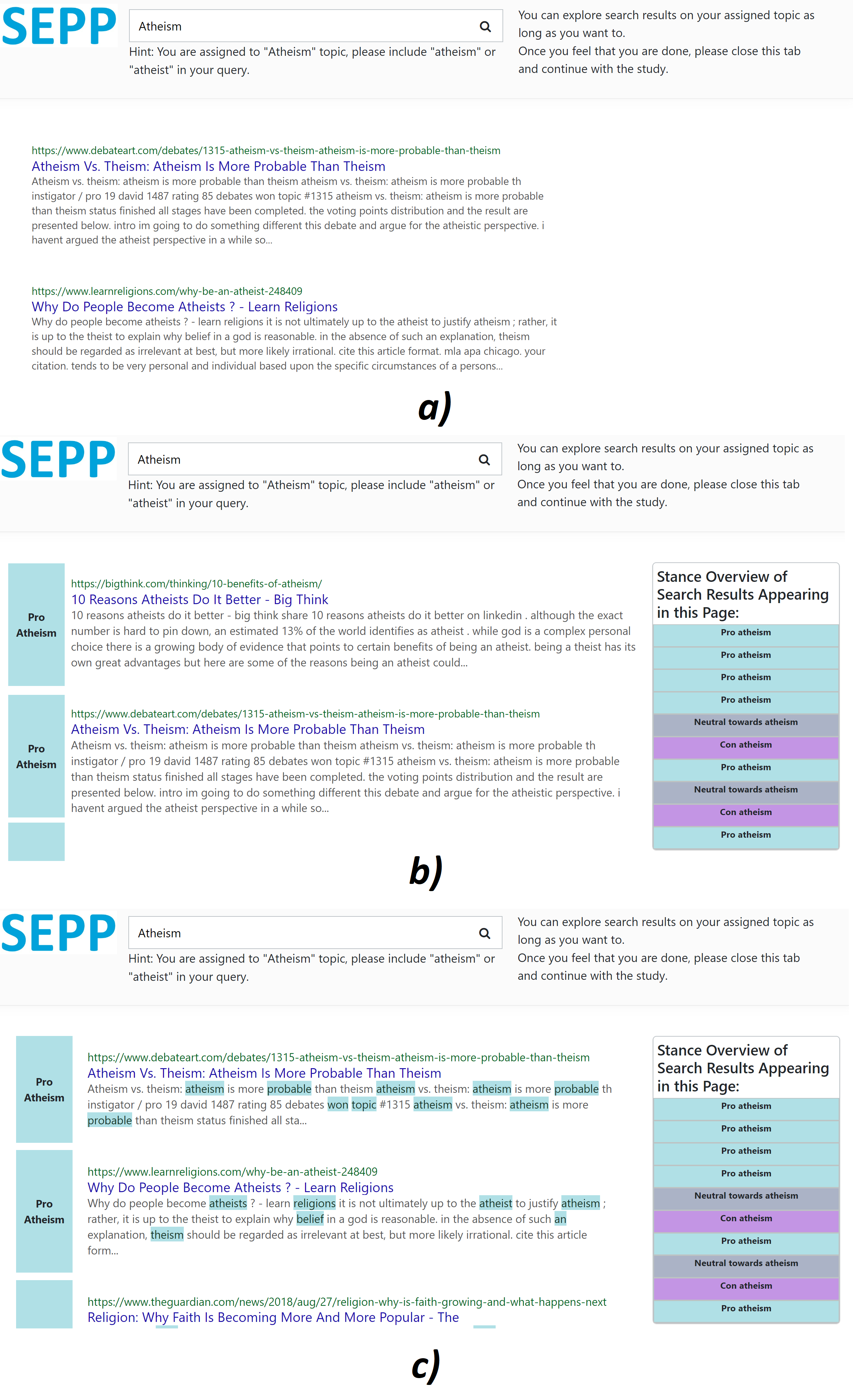}
    \caption{Different conditions for the SERP display. \textit{a)} Text-only Search Result Page. \textit{b)} Predicted stance labels. \textit{c)} Predicted stance labels and Explanations. }
    \label{fig:interface}
\end{figure}

We designed the search interface as follows.
As soon as users open the homepage of the study, saw a search bar positioned at the top of the page, where they may enter their queries. The search engine showed results only when the input query included one or more keywords referring to the user's assigned topic (among atheism, intellectual property rights, and school uniforms). 
Otherwise, showed a message informing users that no results were found for their query. We provided users with the topic and the specific keyword to include in the query with a sentence below the search bar.
For the \textit{Atheism} topic, the keywords are \enquote{atheist} or \enquote{atheism}, for the \textit{intellectual property right} topic, the keywords are \enquote{property right}, \enquote{intellectual right} or \enquote{intellectual property right}, while for the \textit{school uniform} topic, keywords are \enquote{uniform} or \enquote{school uniform}. 

After the users inserted a query including the above mentioned keywords, the interface displayed a list of matched results. This set of search results had a different arrangement based on the random condition a user was assigned to. We set up three SERP display conditions (see Section \ref{sec:variables}): 1) text-only SERP, 2) SERP with labels; and 3) SERP with labels and explanations. The text-only SERP showed ten search results without any extra information. Each search result had two parts: a clickable title redirecting the user to the corresponding webpage and a snippet showing some of its content. Figure \ref{fig:interface} a) shows the layout of this type of interface. The SERP with labels introduces two new features: the labels indicating the original content's stance and an overview of all the search results labels on the right side of the webpage. As shown in Figure \ref{fig:interface} b), we assigned a different color representing each viewpoint. We also added extra information to each label, indicating whether the document is \textit{against}, \textit{neutral}, or \textit{in favor} of the topic. 
The third type of SERP interface makes use of a mono-chromatic saliency map (see Figure \ref{fig:interface} c), highlighting the top 20 words that best contribute 
to the prediction in the search result snippet.
Due to the space limitation of the interface, not all the feature would appear in the snippet.
The color of the saliency map aligns with the color of the label.
We use the same color for all 20 feature words regardless of their significance to make it less complicated for users to understand.

\subsubsection{Biased Versus Balanced Results.}
\begin{table}[h!]
    \caption{Templates for the first SERP (i.e., the top 10-ranked search results) in each SERP ranking bias condition.}
    \label{tab:templates}
    \centering
    \setlength{\tabcolsep}{12pt}
    \begin{tabular}{p{1cm}p{1.8cm}p{2cm}p{2cm}}
        \toprule
        \textbf{Rank} & \textbf{Biased Opp. (T1)} & \textbf{Biased Supp. (T2)} & \textbf{Balanced (T3)} \\
        \midrule
        \small{1} & Against & Favor & Neutral\\
        \small{2} & Against & Favor & Neutral\\
        \small{3} & Against & Favor & Neutral\\
        \small{4} & Against & Favor & Neutral\\
        \small{5} & Favor & Against & Against\\
        \small{6} & Neutral & Neutral & Neutral\\
        \small{7} & Against & Favor & Favor\\
        \small{8} & Favor & Against & Against\\
        \small{9} & Neutral & Neutral & Neutral\\
        \small{10} & Against & Favor & Favor\\
        \bottomrule
    \end{tabular}
\end{table}

We ranked search results depending on the ranking bias condition randomly assigned to each user. Specifically, we created \textit{biased} and \textit{balanced} top 10 search result ranking templates, according to which we would later display search results to users. Biased search results were biased either in the \textit{against} or \textit{in favor} direction and thus contain only results of one of these two stance classes in the top four ranking spots (the remaining six results were balanced across stance classes). Users who were assigned to the \textit{biased} condition would see search results biased in the opposite direction compared to their pre-search opinion (e.g., if they were in favor of school uniforms, they would see results biased against school uniforms). Users who were assigned to the balanced condition would see a similar search result page, but with neutral search results in the top four spots. Neutral results either do not contain any arguments on the debated topic or equally many arguments in both directions.\footnote{We chose this setup to make the conditions as comparable as possible, e.g., rather than displaying results in alternating fashion in the balanced condition.} Table \ref{tab:templates} shows an overview of the ranking templates.


\section{User Study Setup}
\label{sec:survey}

\subsection{Research Ethics and pre-registration.}
We deployed the web application on servers owned by the Faculty of Maastricht University (UM server) 
and secured the connection through HTTPS protocol with SSL certificates. 
The study was reviewed and accepted by our review board. The data collected from users, ranging from demographical information to user's clicking behaviours, are all stored anonymously in the server. 
Prior to the launch of the user study, the research question, hypotheses, methodology, measurements, etc. were pre-registered on the Open Science Framework.
Only minor changes were made, including balancing the number of participants in the different conditions, change the way we measure users' attitude change, and correcting a computation error in the macro F1 score.

\subsection{Variables}
\label{sec:variables}
In our study, each subject could look at search results (i.e., 10 search results per page) accompanied by different features.
We analyzed the participants' attitudes and interaction behavior (with a focus on the proportion of clicks on attitude-confirming search results).

\vspace{.5em}
\noindent \textbf{Independent variables.}

\begin{itemize}[leftmargin=*]
    \item \textit{Topic} (between-subjects, categorical). Participants were assigned to one topic (i.e., atheism, intellectual property rights, or school uniforms) for which they have a strong pre-search attitude (i.e., strongly opposing or strongly supporting). If a participant had no strong attitude on any topic, they ended the study. If a participant had multiple strong attitudes, they were assigned to the topic that has the fewest participants at that point in the study (i.e., to move toward a balanced topic distribution).
    
    \item \textit{SERP ranking bias} (between-subjects, categorical).
    There were two types of ranking conditions: biased and balanced.
    For each of these two conditions, we preset a ranking template (see Table \ref{tab:templates}).
    Participants would see a search result page with ten items which were ranked in accordance with the template. If a user was assigned to the biased condition, they would see opposing-biased search results if their pre-search attitude was \textit{in favor} (i.e., $3$) and supporting-biased search results if their pre-search attitude was \textit{against} (i.e., $-3$).

    \item \textit{SERP display} (between-subjects, categorical). Each participant saw search results accompanied by one of these features: (1) plain text results without stance labels, (2) results with predicted stance labels (Figure~\ref{fig:interface} b), or (3) results with predicted stance labels and highlighted explanations (Figure~\ref{fig:interface} c).

\end{itemize}

\noindent \textbf{Dependent variable.}

\begin{itemize}[leftmargin=*]
    \item \textit{Shannon Index} (numerical). The Shannon Index was applied to measure the diversity of users' clicks.
    Let $N$ be the total number of clicks made in one session, $n_0, n_1, n_2$ be the number of clicks of "against", "neutral" and "favor" items respectively.
    The formula for computing clicking diversity is: $-\sum_{i=0}^{2}\frac{n_i}{N} \ln(\frac{n_i}{N})$ .
    The convention for no occurrence of a class is to ignore it as $\ln(0)$ is undefined~\cite{mackay2003information}. 
    For instance, the Shannon Index for (0, 1, 3) is $0 + 0.34 + 0.21 = 0.55$.
    The minimum value of the Shannon Index is 0, which indicates that there is no diversity and only one viewpoint was clicked on.
    When each class is equal the Shannon entropy has the highest value (for three classes, this would be $3*(-\frac{1}{3})\ln(\frac{1}{3})=1.1$).
\end{itemize}

\vspace{.5em}
\noindent \textbf{Descriptive and exploratory measurements.} We used these variables to describe our sample and for exploratory analyses, but we did not conduct any conclusive hypothesis tests on them.

\begin{itemize}[leftmargin=*]
    \item \textit{Demographics} (categorical). We asked participants to state their gender, age group, and level of education from multiple choices.
    Each of these items includes a ''prefer not to say'' option.
    \item \textit{Clicks on Neutral Items} (numerical). In a balanced SERP, the majority of items were neutral.
    We were specifically interested in whether participants' engagement with search results with a neutral stance is affected by the SERP display condition.
    \item \textit{Clicking Diversity} (numerical). We logged the clicking behavior of participants during the survey and computed the ratio of pre-search attitude-confirming vs. attitude-opposing search results among the results a user has clicked on.
    Clicks on neutral search results were not regarded for this variable.
    \item \textit{Attitude Change} (numerical). In line with previous research~\cite{draws2021ThisNotWhat,rieger2021ThisItemMight,epstein2015SearchEngineManipulation}, we asked participants to select their attitudes on debated topics before and after the experiments using a seven-point Likert scale ranging from ``strongly opposing'' to ``strongly supporting''.
    The difference between their two answers is then assessed in the analysis.
    
    \item \textit{Textual feedback} (free text). We asked participants to provide feedback on the explanations and the task.

\end{itemize}

\subsubsection{Procedure.}
Participants completed the study in three steps as described below. The survey was conducted on Qualtrics\footnote{\url{https://www.qualtrics.com/}}, while the interaction with search results occurred on our own server.

\paragraph{\textbf{Step 1.}}
After agreeing to an informed consent, participants were asked to report their gender, age group, and level of education. 
Participants were first asked to imagine the following scenario: 
\begin{displayquote}
\textit{You and your friend were having a dinner together. Your friend is very passionate about a debated topic and couldn't help sharing his views and ideas with you. After the dinner, you decide to further inform yourself on the topic by conducting a web search.}
\end{displayquote}
Furthermore, participants were asked to state their attitudes concerning each debated topic (see Section \ref{sec:datapreparation}; including one attention check for which we specifically instruct participants on what option to select from a Likert scale).

\paragraph{\textbf{Step 2.}} We introduced participants to the task and subsequently assigned them to one of the three debated topics (i.e., \textit{atheism}, \textit{intellectual property rights}, and \textit{school uniforms}) depending on their pre-search attitudes and randomly assigned them to one SERP ranking bias condition and one SERP display condition (see Section \ref{sec:variables}).
Participants were then asked to click on a link leading them to our search platform (i.e., SEPP; see Section \ref{sec:sepp}). Here, participants could enter as many queries as they want, as long as those queries include their assigned topic term (e.g., \texttt{school uniforms pros and cons} for the topic \textit{school uniforms}).
Regardless of what or how many queries participants enter, they always received search results from the same pool of 21 available search results relevant to their assigned topic (i.e., seven against, seven neutral, and seven in favor; see Section \ref{sec:datapreparation}).
With every query that participants entered, they received those search results ranked according to the ranking template associated to their assigned SERP ranking bias condition (see Section \ref{sec:sepp}) for the first SERP and randomly drawn search results (following the template) for consequent searches.\footnote{Whenever a user enters a new query, the first SERP (i.e., displaying the top 10 results) will always show search results according to the template, whereas pages 2 and 3 will show the 21 search results relevant to the topic in random order.}
Depending on the SERP display condition participants were assigned to, they could see either plain search results, search results accompanied by stance labels, or search results accompanied by stance labels with additional explanations.
Participants were made aware that the search results they were seeing might be biased and that there were limited results.
After entering a query, participants were free to explore search results as long as they wish and click on links that lead to the presented web pages, or enter new queries.
Users were instructed to return to the Qualtrics survey when they were done searching.

\paragraph{\textbf{Step 3.}} Finally, in the questionnaire, we asked participants to report their post-search attitude (towards their assigned topic). 
Further, we asked them to provide textual feedback on the explanations and the task.
We also included another attention check to filter out low-quality data in this post-interaction questionnaire. 
The attention checks consisted of one straightforward question with suggested response options. 
We excluded the data of participants who failed one or more of the attention checks from data analysis. 

\subsubsection{Recruitment Methods.}
In this study, we used Prolific\footnote{\url{https://prolific.co}} and Qualtrics\footnote{\url{https://www.qualtrics.com/}} to manage the participants and design survey workflow, respectively.
The workflow of our study, including the informed consent, screening questions, and link to the survey, were all completed on Qualtrics.
In the recruitment platform Prolific, we only selected participants with a minimum age of 18 (in compliance with research ethics), and fluent in English, as our dataset only contains English results.

\subsubsection{Sample Details.}
We anticipated to observe medium effects for \textit{SERP display} and \textit{SERP ranking bias} on \textit{clicking diversity} (Cohen's $f = 0.25$). 
Thus, we determined in an a priori power analysis for a between-subjects ANOVA (see Section~\ref{sec:statAnalyses}) a required sample size of 205 participants, assuming a significance threshold of \textalpha~= $\frac{0.05}{3}$ = 0.017 (testing three hypotheses), a desired power of (1-~\textbeta) = 0.8 and considering that we tested, depending on the hypothesis, six groups (i.e., three SERP display conditions: \textit{without stance labels, with stance labels, with stance labels and explanation}; and 2 SERP ranking bias conditions: \textit{biased towards the attitude-opposing viewpoint, 
balanced}) using the software \textit{G*Power} \cite{faul2009statistical}. 
We aimed for a balanced distribution across topics and conditions. 

Participants were required to be older than 18 and with a high proficiency of English (i.e., as reported by \textit{Prolific}). Participants could only participate in our study once.
As mentioned above, we excluded participants from data analysis if they did not pass one or more attention checks. 
We also excluded participants from data analysis if they did not access our search platform at all or if they did not click on any links during their search.

\subsubsection{Statistical Analyses.}
\label{sec:statAnalyses}
To test our three hypotheses, we conducted Analysis of Variance (ANOVA), looking at the main and interaction effects of the three independent variables (1) the topic, (2) \textit{SERP display} (without stance labels, withstance labels, with stance labels and explanation) and (3) \textit{SERP ranking bias} (biased towards the attitude-opposing viewpoint, 
balanced) on the \textit{shannon index} (H1a, H1b, H1c).
Aiming at a type 1 error probability of $\alpha = 0.05$ and applying Bonferroni correction to correct for multiple testing, we set the significance threshold to $\frac{0.05}{3}=0.017$.
We added \textit{topic} as an additional independent variable to this analysis to control for its potential role as a confounding factor.

In addition to the analyses described above, we conducted posthoc tests (i.e., to analyze pairwise differences) to determine the exact differences and effect size, Bayesian hypothesis tests (i.e., to quantify evidence in favor of null hypotheses), and exploratory analyses (i.e., to note any unforeseen trends in the data) to better understand our results.


\section{Results}
The overarching objective of this research is to investigate the effect of extra visual elements such as stance labels and text explanations on users' interaction behaviours, especially in terms of clicking diversity.
In this section, we present the final results of the study with 203 participants and address our research question and hypotheses.

\subsubsection{Descriptive Statistics.}
Prior to analyzing the primary statistics, it is necessary to first examine the demographic data.
\begin{figure}[!ht]
    \centering
    \includegraphics[width=0.9\textwidth]{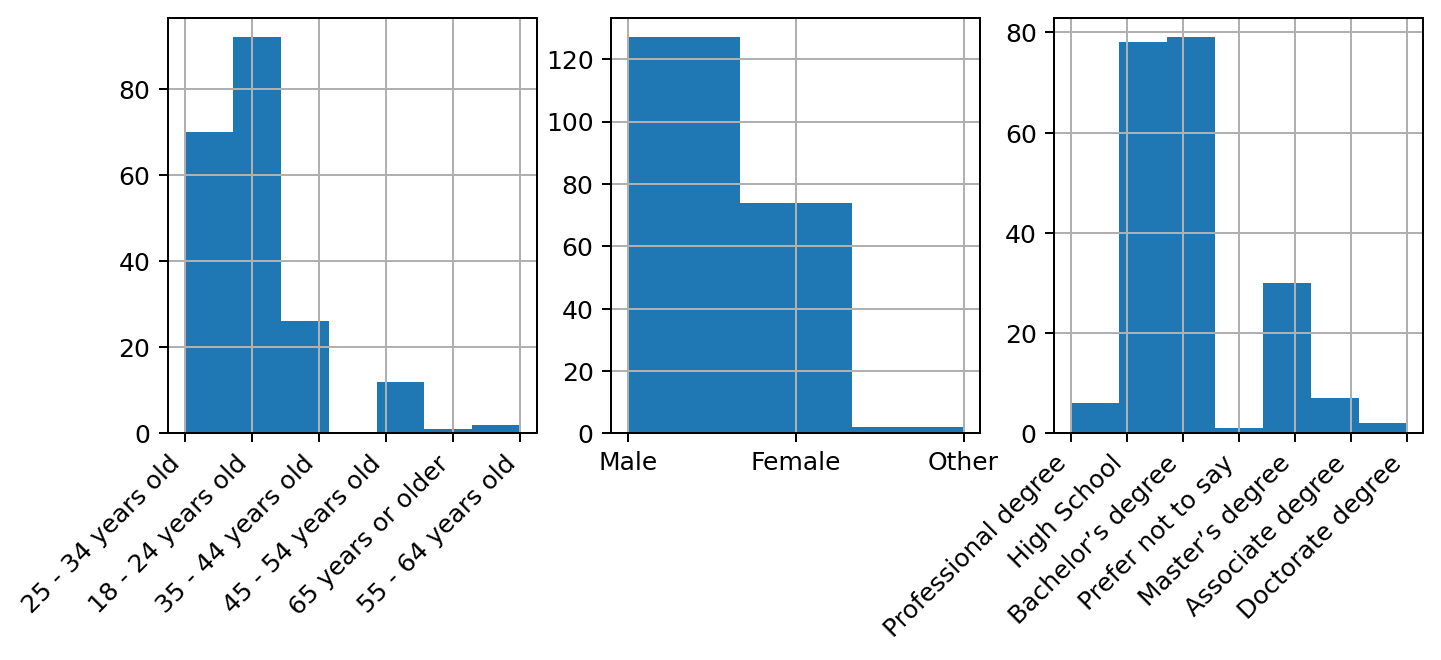}
    \caption{Demographic information of participants: Bar chart showing the distribution of users' age, gender, and education level.}
    \label{fig:participantDist}
\end{figure}
In general, young people made up most of our participants (see Figure \ref{fig:participantDist}).
The educational level data reveal that the majority of participants have completed at least some level of higher education, with a smaller percentage have completed advanced degrees.

Participants were roughly equally distributed across the three topics: 70 \textit{atheism}, 66 \textit{intellectual property rights}, and 67 \textit{school uniforms}.                 
Regarding factors such as bias and interface types that were randomized by Qualtrics workflow, their distributions are also balanced, with 102 participants accessed the biased SERPs and 101 accessed balanced SERPs, and 72 users view text-only SERP display, 64 viewed labelled interface and 67 viewed interface with saliency maps.
In the pre-survey attitude test, 130 people were granted the access to our survey by expressing rather negative (against) viewpoints towards a specific topic, while only 73 people expressed a positive stance.

\begin{figure}[ht]
    \centering
    \includegraphics[width=0.8\textwidth]{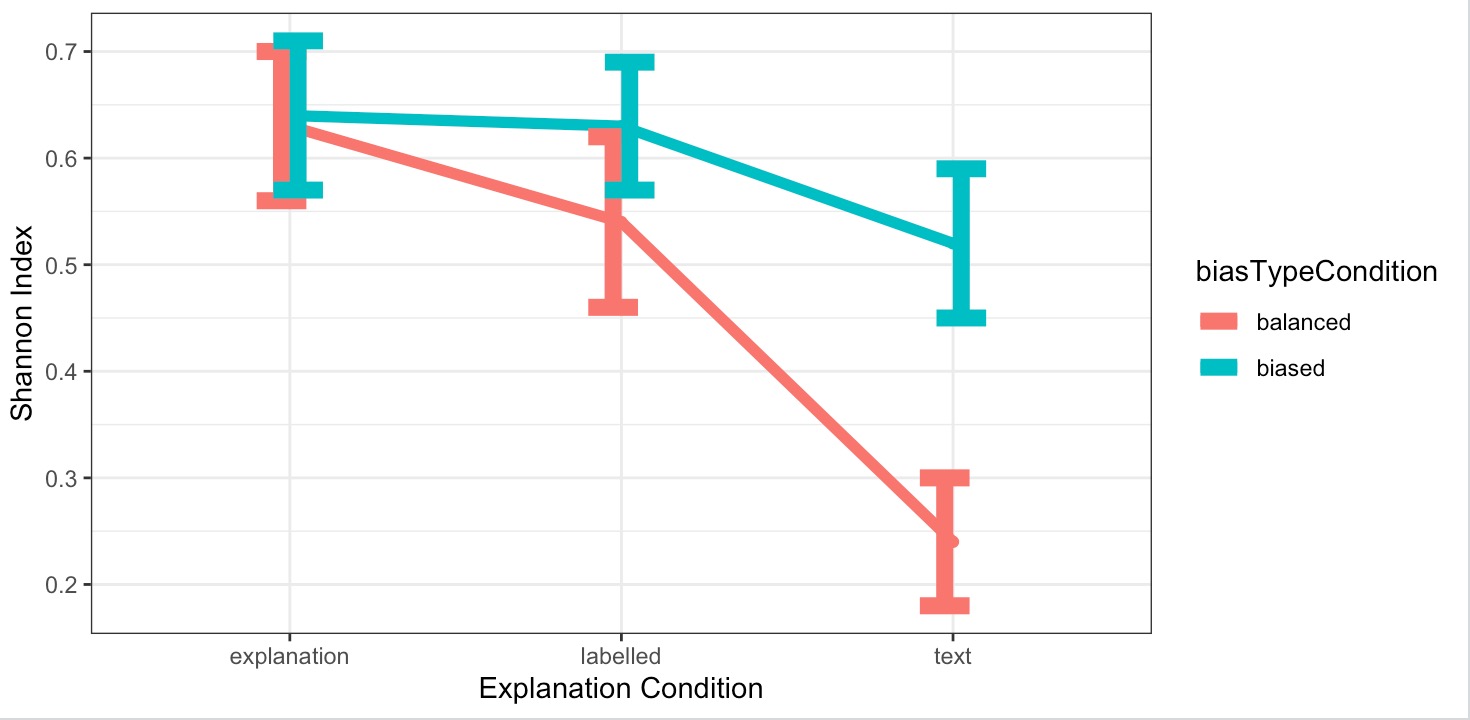}
    \caption{
    Shannon Index across SERP Display conditions, split by ranking bias conditions. Error bars represent confidence intervals.} 
    \label{fig:userstudy}
\end{figure}

Figure \ref{fig:userstudy} shows the diversity of users' clicks as measured by the Shannon index across conditions. 
For balanced SERPs, the mean Shannon diversity index starts from 0.63 when there are explanation, then it slightly reduces to 0.55 in labelled interface, and drastically drop to 0.24 for the text SERP.
For unbalanced SERPs, the trend is similar, from 0.64 to roughly 0.52, but the reduction is less drastic.
In short, those who interacted with balanced pages overall scored somewhat lower in Shannon diversity. 

\begin{figure}[ht]
    \centering
    \includegraphics[width=0.8\textwidth]{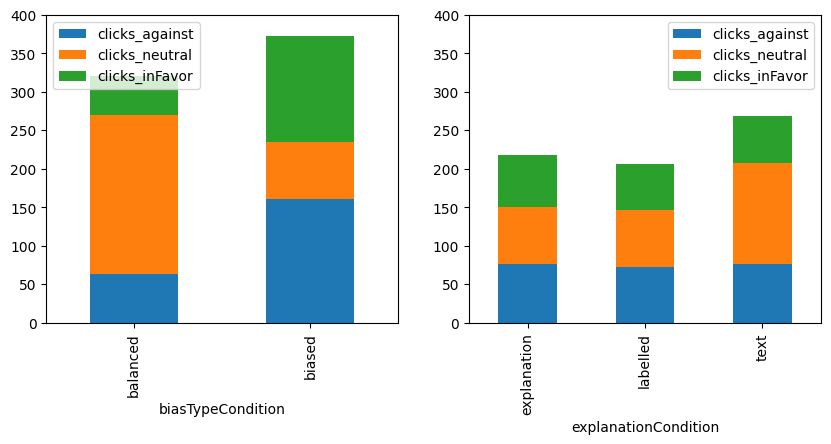}
    \caption{Number of clicks on items of each stance made by users under different conditions. Left: different bias type, Right: different interface}
    \label{fig:clickss}
\end{figure}

Figure \ref{fig:clickss} shows the stacked histogram representing users' click history under different conditions.
In the left subplot, we see that \textit{neutral} items attracted more clicks than \textit{favor} and \textit{aginst} combined in the balanced setting.
In the biased setting, however, the number neutral clicks starts to shrink and user start to visit polarized contents more.
It is notable that users clicked more \textit{neutral} items in the text-only SERP display. The other two types of interface seem to generate very similar amounts of clicks for each stance.

\subsection{Hypothesis Tests}
We ran an ANOVA test to examine the relationship between the variable \textit{Shannon index} and other predictor variables, including \textit{explanation condition}, \textit{bias type} and \textit{assigned topic}.
Table \ref{tab:anova} contains the results of the ANOVA test, including the F-statistic, p-value, and degrees of freedom for each variable and interaction.
We also included the assigned topic in the test to control for topic as a potential confounding factor. 

\begin{table}[!ht]
  \centering
  \setlength{\tabcolsep}{7pt}
    \begin{tabular}{p{3.5cm}ccccc}
    \toprule
    \textbf{Variable} & \textbf{Df} & \textbf{Sum Sq} & \textbf{Mean Sq} & \textbf{F} & \textbf{$p$} \\
    \midrule
    SERP Ranking Bias  & 1     & 0.81  & 0.8146 & 4.911 & .027  \\
    SERP Display & 2     & 2.54  & 1.2712 & 7.664 & $<$.001 \\
    Topic & 2     & 0.03  & 0.0158 & 0.095 & .909 \\
    SERP RB:SERP Display & 2     & 0.65  & 0.3256 & 1.963 & .143 \\
    Residuals & 195   & 32.34 & 0.1659 &       &  \\
    
    \bottomrule
    \end{tabular}%
  \caption{ANOVA results for the Shannon index and independent variables including the explanation level, bias type, the interaction between explanation and bias type, and the topic the user holds strong opinion towards. 
  Under our reduced significance threshold, only the \textit{Explanation} effect is significant.
  }
  \label{tab:anova}%

\end{table}%

\subsubsection{H1a: Users who are exposed to viewpoint-biased search results interact with less diverse results than users who are exposed to 
balanced search results.}

To test this hypothesis, we examine the Shannon diversity of clicks made by users who viewed viewpoint-biased search engine result pages (SERPs) to those who did not. Figure \ref{fig:userstudy} suggests that users who were exposed to more balanced search results clicked on somewhat less diverse content. 
However, in the ANOVA summary (Table \ref{tab:anova}), we see that the influence of bias type ($F = 4.911, df = 1, p = .027$, Cohen's $f = 0.16$) is not statistically significant, given the significance threshold of .017. We thus do not find any conclusive evidence for a difference in the diversity of clicked results between bias conditions.

\subsubsection{H1b: Users who are exposed to search results with (1) stance labels or (2) stance labels with explanations for each search result interact with more diverse content than users who are exposed to regular search results.}

Regarding H1b, compare users' clicks in different SERP displays.
The results of the ANOVA shows a significant effect of explanation condition on clicking diversity ($F = 7.664, df = 2, p < .001$, Cohen's $f = 0.28$). 
This suggests that there is a difference in the diversity of content interacted between users who were exposed to (1) plain search results, (2) search results with stance labels, or (3) search results with stance labels and also explanations.
We conducted a pair-wise Tukey test to determine whether there are significant differences between SERP display levels (text, label, explanation). We found significant differences between \texttt{text-label} and \texttt{text-explanation} (with adjusted p-values of .0008 and .012, respectively), suggesting that both stance labels and stance labels with explanations led to more diverse clicks compared to the regular SERPs.
The p-value of \texttt{labelled-explanation} group is .723, indicating that there may be no difference in Shannon diversity between the \textit{labelled} and \textit{explanation} group.

\subsubsection{H1c: Users who are exposed to search results with (1) stance labels or (2) stance labels with explanations are less susceptible to the effect of viewpoint biases in search results on clicking diversity.}

This hypothesis concerns the interaction effect between SERP display and bias types on the Shannon index.
Our ANOVA (Table \ref{tab:anova}) did not reveal any evidence for such an interaction effect ($F=1.963, df = 2, p=.143$, Cohen's $f = 0.02$).
In other words, when users are looking at different types of SERP displays but are exposed to the same level of viewpoint bias, our results do not contain evidence that users will click more diverse items because explanations and labels are visible.

\subsection{Exploratory Analysis}
To further understand our results, we also conducted exploratory analysis. Note that these analyses were not preregistered.

\subsubsection{Clicks on Neutral Items.}
To better understand why the search results were more diverse in the labeled and explanation conditions, we looked closer at the distribution across the three viewpoints. 
In Figure \ref{fig:clickss}, we can already see a larger number of neutral results for the text-only condition. 
Furthermore, in every SERP display, users who were exposed to a balanced page clicked on more neutral items on average.
This may be due to their tendency to click on highly-ranked neutral items while being unaware of the stance.

We conducted an exploratory ANOVA to investigate the effects of biases and interfaces on the number of clicks on neutral items.
Table \ref{tab:neutralanova} lists the test results.
\begin{table}[!ht]
  \centering
  \setlength{\tabcolsep}{7pt}
    \begin{tabular}{p{3.5cm}ccccc}
    \toprule
    \textbf{Variable} & \textbf{Df} & \textbf{Sum Sq} & \textbf{Mean Sq} & \textbf{F} & \textbf{$p$} \\
    \midrule
    SERP Ranking Bias  & 1     & 84.3  & 84.30 & 44.308 & $<$.001  \\
    SERP Display  & 2     & 23.1  & 11.54 & 6.067 & .003  \\
    SERP RB:SERP Display & 2     & 15.6  & 7.79  & 4.096 & .018  \\
    Residuals & 197   & 374.8 & 1.90  &       &  \\
    \bottomrule
    \end{tabular}%
    \caption{Results of the ANOVA analysis on for number of neutral clicks. 
    }
  \label{tab:neutralanova}

\end{table}%
We can observe that both explanation condition and bias condition had main effects on click diversity.
Also their interaction is significant. This means that users may click on more neutral results in the balanced condition, and especially so when SERPs do not contain any stance labels or explanations.

\subsubsection{Attitude Change.}
\begin{figure}[!ht]
    \centering
    \includegraphics[width=0.9\textwidth]{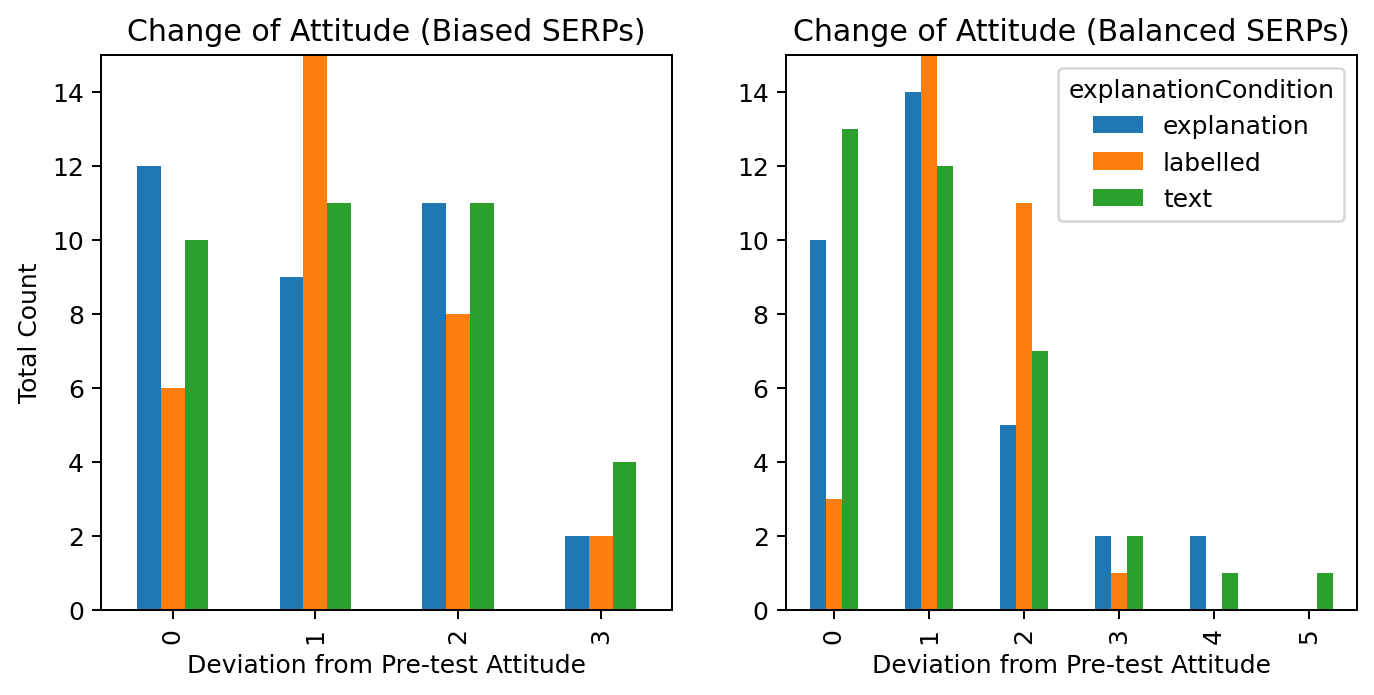}
    \caption{
    Histogram visualizing the difference between pre- and post-survey attitudes. 
    Each group of bins indicates the number of people in a bias type setting who changed their viewpoints towards their assigned topic by a certain number of points.
    The X-axis indicates the normalized difference of attitude between the pre- and post-survey answers measured by a 7-point Likert Scale.
    }
    
    \label{fig:attidutechange}
\end{figure}

Previous research showed that mildly opinionated users' attitude change can differ across levels of ranking bias~\cite{draws2021ThisNotWhat,epstein_search_2015,pogacar2017PositiveNegativeInfluence,allam2014ImpactSearchEngine,bink2023InvestigatingInfluenceFeatured}.
However, this effect has, to the best of our knowledge, not yet been shown for strongly opinionated users such as in our study.
We intended to measure the attitude change by subtracting the post-search viewpoint $t_{1}$ from the pre-search viewpoint $t_{0}$, and thus the difference would range from $-6$ to $6$.
We summarize users' attitude change in Figure \ref{fig:attidutechange} and Table \ref{tab:attitudechange}.

\begin{table}[!ht]
\scriptsize
    \centering
    \setlength{\tabcolsep}{3pt}
    \begin{tabular}{c|c|c|c|c}
    \toprule
        \textbf{Bias Type} & \textbf{SERP display} & 
        \textbf{Attitude Change} &  \textbf{Attitude Change} & \textbf{Skewness} \\
        & & (Mean) & (Median) & \\
        \midrule
        \multirow{3}{4em}{Balanced} & Text-only & 1.14 & 1.00 & 1.31\\ 
            & Labelled &  1.30 & 1.00 & 0.15 \\ 
            & Explanation &  1.15 & 1.00 & 1.10\\ 
        \midrule
        \multirow{3}{4em}{Biased} & Text-only & 1.25  & 1.00 & 0.20\\ 
            & Labelled &  1.19 & 1.00 & 0.35\\ 
            & Explanation &  1.09 & 1.00 & 0.25\\ 
\bottomrule
    \end{tabular}
    \caption{Absolute values of attitude change per condition.}
    \label{tab:attitudechange}
\end{table}

From Figure \ref{fig:attidutechange}, we observe that only a few participants developed large attitude changes (absolute value $>$ 3) after the survey. In Table \ref{tab:attitudechange}, the mean absolute attitude change ranges from 1.0 to 1.3. 
The median statistics appear to be very stable; under all conditions, 1 is always the central value.
We performed an exploratory ANOVA on the absolute \textit{attitude change} variable (Table \ref{tab:anovaattidutechange}) but do not find any convincing evidence for an effect of any independent variables on attitude change in our scenario, i.e., apart from a potential role of the topic ($F=4.132,df=2,p=.017$) that would need to be further investigated.

\begin{table}[!ht]
\centering
\setlength{\tabcolsep}{7pt}
\begin{tabular}{lccccc}
\toprule
\textbf{Variable} & \textbf{Df} & \textbf{Sum Sq} & \textbf{Mean Sq} & \textbf{F} & \textbf{$p$}\\
\midrule
SERP Ranking Bias & 1 & 0.01 & 0.013 & 0.014 &.907 \\
SERP Display & 2 & 0.56 & 0.282 & 0.296 & .744 \\
Topic & 2 & 7.88 & 3.938 & 4.132 & .017 \\
SERP RB:SERP Display & 2 & 0.55 & 0.277 & 0.290 & .748 \\
Residuals & 195   & 185.88 & 0.953  &       &  \\

\bottomrule
\end{tabular}
\caption{ANOVA Results: absolute value of users' attitude change}
    \label{tab:anovaattidutechange}

\end{table}

\subsection{Qualitative feedback}
After excluding feedback with fewer than four characters like “no”, “nope” etc, there are 59 substantial comments.
Positive feedback indicated that participants perceived the search results to be diverse: \textit{“I thought the search engine was quite varied as it provided different types of sources”}.
Other participants perceived the viewpoint labels to be accurate: \textit{“The search engine was very useful, and the classification of the information was accurate.”}
At least some participants were able to perceive the bias in the search result pages: \textit{“the results against atheism were promoted to the top results of the search, it was not an impartial search result”.}
While we aimed to mimic a real search engine, some participants may have realized that the results were reshuffled:\textit{ “It appears the results of the search were similar or the same in each new search. I'm not sure the words used to determine the stance of the page were appropriate.”}

\section{Discussion and Limitations} \label{sec:discussion}

We found a significant effect of \textit{SERP display} on clicking diversity. That is, participants who were exposed to viewpoint labels or viewpoints labels with explanations 
consumed more diverse results than plain text search. 
While both non-text SERP displays affect users' click diversity over text-only SERP displays, we cannot observe any additional effect from the SERP display with labels and explanations over the label-only SERP display.
Our results suggest that this difference can be explained by a predominance of clicks on the neutral stance in the text-only condition. This, alongside the qualitative comments, suggests that participants trusted and used the labels and explanations to inform themselves diversely. Contrary to our expectations, we did not find evidence for a difference between \textit{biased and balanced search results}, in terms of click diversity. 
Furthermore, exploratory analyses revealed that users exposed to balanced SERPs clicked on more neutral items.

We further found no exploratory evidence that intervention types (bias of search results; explanation level) affect participants' viewpoints. This later result is not necessarily surprising, given that the participants in this study held strong opinions on the topic. We also would not expect a change because the task given to participants was formulated as low-stakes and open-ended (inform yourself about the topic after speaking to an opinionated friend).

\textbf{Limitations and Future work.} Our study has at least two important limitations. First, each search result in the data set we considered only had one overall viewpoint label, and this was limited to \textit{against}, \textit{neutral}, and \textit{in favor}. This allowed for scalability but does not reflect the full nuances of online search results. For example, an essay or blog post could express highly diverse perspectives but still receive a positive stance label if that is its overall conclusion.  
Secondly, our study carefully aimed to balance a controlled environment with maintaining ecological validity. Despite this, it is possible that participants recognized that this was not a true search engine when issuing new queries.
Similarly, the selection of templates allowed for some structured reshuffling of results. However, this also meant that strong opinions (contrary to the active user) were always at the top of biased search results. For the balanced condition there were also more neutral results. This likely contributed to the higher diversity of search results selected in the biased condition. In addition, ranking fairness metrics could have been used for the ranked lists, which could have led to slightly different results.
Further work is required to disentangle the relationship between position bias and confirmation bias, and to replicate the study with different templates.

\section{Conclusion}
In this paper, we studied the impact of stance labels and explanations of these labels for search on disputed topics. 
We found that stance labels and explanations led to a more diverse search result consumption.
However, we cannot conclude that explanations have an extra effect in addition to labels on user's click diversity. Whether consuming diverse or more neutral results is preferable is in itself a debated topic. 
Backfire effects, where users become even more invested in their pre-existing beliefs, are possible when users consume strongly opposing views~\cite{nyhan2010corrections}. Greater diversity can further induce wrong perceptions of present evidence: for example, portraying climate change deniers and believers equally can give the impression that climate change is an open issue and thus may be worse than indeed weighing the evidence on both sides~\cite{cushion2019quantitative}. How much stance diversity is ideal can thus depend on individual user traits~\cite{munson2010presenting} and may lie somewhere between the extremes~\cite{bail2018exposure}. 
While these are domains and a context where confirmation biases are expected to be large, similar cognitive biases are likely to occur in other decision-making tasks where XAI is used. Further work is needed to catalogue different scenarios of opinion formation in different domains where confirmation bias may be present.

 \section*{Acknowledgments}
     This project has received funding from the European Union’s Horizon 2020 research and innovation programme under the Marie Skłodowska-Curie grant agreement No 860621. 

\newpage 

\bibliographystyle{splncs04}
\bibliography{xai1.bib}

\begin{thebibliography}{10}
\providecommand{\url}[1]{\texttt{#1}}
\providecommand{\urlprefix}{URL }
\providecommand{\doi}[1]{https://doi.org/#1}

\bibitem{aldayel2019your}
Aldayel, A., Magdy, W.: Your stance is exposed! analysing possible factors for stance detection on social media. Proceedings of the ACM on Human-Computer Interaction  \textbf{3}(CSCW),  1--20 (2019)

\bibitem{aldayel2021StanceDetectionSocial}
Aldayel, A., Magdy, W.: Stance detection on social media: State of the art and trends. Information Processing \& Management  \textbf{58}(4),  102597 (Jul 2021). \doi{10.1016/j.ipm.2021.102597}

\bibitem{allam2014ImpactSearchEngine}
Allam, A., Schulz, P.J., Nakamoto, K.: The {Impact} of {Search} {Engine} {Selection} and {Sorting} {Criteria} on {Vaccination} {Beliefs} and {Attitudes}: {Two} {Experiments} {Manipulating} {Google} {Output}. Journal of Medical Internet Research  \textbf{16}(4), ~e100 (Apr 2014). \doi{10.2196/jmir.2642}, \url{http://www.jmir.org/2014/4/e100/}

\bibitem{allaway-mckeown-2020-zero}
Allaway, E., McKeown, K.: {Z}ero-{S}hot {S}tance {D}etection: {A} {D}ataset and {M}odel using {G}eneralized {T}opic {R}epresentations. In: Proceedings of the 2020 Conference on Empirical Methods in Natural Language Processing (EMNLP). pp. 8913--8931. Association for Computational Linguistics, Online (Nov 2020). \doi{10.18653/v1/2020.emnlp-main.717}, \url{https://aclanthology.org/2020.emnlp-main.717}

\bibitem{augenstein2016StanceDetectionBidirectional}
Augenstein, I., Rockt{\"a}schel, T., Vlachos, A., Bontcheva, K.: Stance detection with bidirectional conditional encoding. In: Proceedings of the 2016 Conference on Empirical Methods in Natural Language Processing. pp. 876--885. Association for Computing Machinery, New York, NY, USA (2016)

\bibitem{azzopardi2021}
Azzopardi, L.: Cognitive biases in search: a review and reflection of cognitive biases in information retrieval  (2021)

\bibitem{bail2018exposure}
Bail, C.A., Argyle, L.P., Brown, T.W., Bumpus, J.P., Chen, H., Hunzaker, M.F., Lee, J., Mann, M., Merhout, F., Volfovsky, A.: Exposure to opposing views on social media can increase political polarization. Proceedings of the National Academy of Sciences  \textbf{115}(37),  9216--9221 (2018)

\bibitem{bink2023InvestigatingInfluenceFeatured}
Bink, M., Schwarz, S., Draws, T., Elsweiler, D.: Investigating the influence of featured snippets on user attitudes. In: ACM SIGIR Conference on Human Information Interaction and Retrieval. CHIIR '23, ACM, New York, NY, USA (2023). \doi{10.1145/3576840.3578323}

\bibitem{bink2022FeaturedSnippetsTheir}
Bink, M., Zimmerman, S., Elsweiler, D.: Featured snippets and their influence on users' credibility judgements. In: ACM SIGIR Conference on Human Information Interaction and Retrieval. pp. 113--122. ACM, Regensburg Germany (Mar 2022). \doi{10.1145/3498366.3505766}

\bibitem{chen2023}
Chen, S., Xiao, L., Kumar, A.: Spread of misinformation on social media: {{What}} contributes to it and how to combat it. Computers in Human Behavior  \textbf{141},  107643 (2023). \doi{10.1016/j.chb.2022.107643}

\bibitem{cushion2019quantitative}
Cushion, S., Thomas, R.: From quantitative precision to qualitative judgements: Professional perspectives about the impartiality of television news during the 2015 uk general election. Journalism  \textbf{20}(3),  392--409 (2019)

\bibitem{draws2023explainable}
Draws, T., Ramamurthy, K.N., Soares, I.B., Dhurandhar, A., Padhi, I., Timmermans, B., Tintarev, N.: Explainable cross-topic stance detection for search results. In: CHIIR (2023)

\bibitem{draws2023ViewpointDiversitySearch}
Draws, T., Roy, N., Inel, O., Rieger, A., Hada, R., Yalcin, M.O., Timmermans, B., Tintarev, N.: Viewpoint diversity in search results. In: Kamps, J., Goeuriot, L., Crestani, F., Maistro, M., Joho, H., Davis, B., Gurrin, C., Kruschwitz, U., Caputo, A. (eds.) Advances in Information Retrieval. pp. 279--297. Springer Nature Switzerland, Cham (2023)

\bibitem{draws2023viewpointdiversity}
Draws, T., Roy, N., Inel, O., Rieger, A., Hada, R., Yalcin, M.O., Timmermans, B., Tintarev, N.: Viewpoint diversity in search results. In: European Conference on Information Retrieval. Springer (2023)

\bibitem{draws2021ThisNotWhat}
Draws, T., Tintarev, N., Gadiraju, U., Bozzon, A., Timmermans, B.: This {Is} {Not} {What} {We} {Ordered}: {Exploring} {Why} {Biased} {Search} {Result} {Rankings} {Affect} {User} {Attitudes} on {Debated} {Topics}. In: Proceedings of the 44th {International} {ACM} {SIGIR} {Conference} on {Research} and {Development} in {Information} {Retrieval}. pp. 295--305. {SIGIR} '21, Association for Computing Machinery, New York, NY, USA (Jul 2021). \doi{10.1145/3404835.3462851}, \url{https://dl.acm.org/doi/10.1145/3404835.3462851}

\bibitem{epstein2015SearchEngineManipulation}
Epstein, R., Robertson, R.E.: The search engine manipulation effect (seme) and its possible impact on the outcomes of elections. Proceedings of the National Academy of Sciences  \textbf{112}(33),  E4512--E4521 (Aug 2015). \doi{10.1073/pnas.1419828112}

\bibitem{epstein_search_2015}
Epstein, R., Robertson, R.E.: The search engine manipulation effect ({SEME}) and its possible impact on the outcomes of elections. Proceedings of the National Academy of Sciences  \textbf{112}(33),  E4512--E4521 (Aug 2015). \doi{10.1073/pnas.1419828112}, \url{http://www.pnas.org/lookup/doi/10.1073/pnas.1419828112}

\bibitem{faul2009statistical}
Faul, F., Erdfelder, E., Buchner, A., Lang, A.G.: Statistical power analyses using g* power 3.1: Tests for correlation and regression analyses. Behavior research methods  \textbf{41}(4),  1149--1160 (2009)

\bibitem{feldhus2022constructing}
Feldhus, N., Hennig, L., Nasert, M.D., Ebert, C., Schwarzenberg, R., M{\"o}ller, S.: Constructing natural language explanations via saliency map verbalization. arXiv preprint arXiv:2210.07222  (2022)

\bibitem{gezici2021EvaluationMetricsMeasuring}
Gezici, G., Lipani, A., Saygin, Y., Yilmaz, E.: Evaluation metrics for measuring bias in search engine results. Information Retrieval Journal  \textbf{24}(2),  85--113 (Apr 2021). \doi{10.1007/s10791-020-09386-w}

\bibitem{gohel2021explainable}
Gohel, P., Singh, P., Mohanty, M.: Explainable ai: current status and future directions. arXiv preprint arXiv:2107.07045  (2021)

\bibitem{hanselowski-etal-2018-retrospective}
Hanselowski, A., PVS, A., Schiller, B., Caspelherr, F., Chaudhuri, D., Meyer, C.M., Gurevych, I.: A retrospective analysis of the fake news challenge stance-detection task. In: Proceedings of the 27th International Conference on Computational Linguistics. pp. 1859--1874. Association for Computational Linguistics, Santa Fe, New Mexico, USA (Aug 2018), \url{https://aclanthology.org/C18-1158}

\bibitem{hardalov2022few}
Hardalov, M., Arora, A., Nakov, P., Augenstein, I.: Few-shot cross-lingual stance detection with sentiment-based pre-training. In: Proceedings of the AAAI Conference on Artificial Intelligence. vol.~36, pp. 10729--10737. AAAI (2022)

\bibitem{jin2019bridging}
Jin, W., Carpendale, S., Hamarneh, G., Gromala, D.: Bridging ai developers and end users: An end-user-centred explainable ai taxonomy and visual vocabularies. Proceedings of the IEEE Visualization, Vancouver, BC, Canada pp. 20--25 (2019)

\bibitem{kaiser2021}
Kaiser, B., Wei, J., Lucherini, E., Lee, K., Matias, J.N., Mayer, J.: Adapting security warnings to counter online disinformation. In: 30th USENIX Security Symposium (USENIX Security 21). pp. 1163--1180 (2021)

\bibitem{kucuk2021StanceDetectionSurvey}
K{\"u}{\c c}{\"u}k, D., Can, F.: Stance detection: A survey. ACM Computing Surveys  \textbf{53}(1),  1--37 (Jan 2021). \doi{10.1145/3369026}

\bibitem{text_rank2023}
Leonhardt, J., Rudra, K., Anand, A.: Extractive explanations for interpretable text ranking. ACM Trans. Inf. Syst.  \textbf{41}(4) (mar 2023). \doi{10.1145/3576924}, \url{https://doi.org/10.1145/3576924}

\bibitem{Lyu2023list}
Lyu, L., Anand, A.: Listwise explanations for\&nbsp;ranking models using multiple explainers. In: Advances in Information Retrieval: 45th European Conference on Information Retrieval, ECIR 2023, Dublin, Ireland, April 2–6, 2023, Proceedings, Part I. p. 653–668. Springer-Verlag, Berlin, Heidelberg (2023). \doi{10.1007/978-3-031-28244-7\_41}, \url{https://doi.org/10.1007/978-3-031-28244-7_41}

\bibitem{mackay2003information}
MacKay, D.J., Mac~Kay, D.J., et~al.: Information theory, inference and learning algorithms. Cambridge university press (2003)

\bibitem{madsen2022post}
Madsen, A., Reddy, S., Chandar, S.: Post-hoc interpretability for neural nlp: A survey. ACM Computing Surveys  \textbf{55}(8),  1--42 (2022)

\bibitem{mena2020}
Mena, P.: Cleaning {{Up Social Media}}: {{The Effect}} of {{Warning Labels}} on {{Likelihood}} of {{Sharing False News}} on {{Facebook}}. Policy \& Internet  \textbf{12},  165--183 (2020). \doi{10.1002/poi3.214}

\bibitem{munson2010presenting}
Munson, S.A., Resnick, P.: Presenting diverse political opinions: how and how much. In: Proceedings of the SIGCHI conference on human factors in computing systems. pp. 1457--1466 (2010)

\bibitem{nickerson1998}
Nickerson, R.S.: Confirmation bias: A ubiquitous phenomenon in many guises p.~46 (1998)

\bibitem{nyhan2010corrections}
Nyhan, B., Reifler, J.: When corrections fail: The persistence of political misperceptions. Political Behavior  \textbf{32}(2),  303--330 (2010)

\bibitem{pogacar2017PositiveNegativeInfluence}
Pogacar, F.A., Ghenai, A., Smucker, M.D., Clarke, C.L.: The positive and negative influence of search results on people's decisions about the efficacy of medical treatments. In: Proceedings of the ACM SIGIR International Conference on Theory of Information Retrieval. pp. 209--216. ACM, Amsterdam The Netherlands (Oct 2017). \doi{10.1145/3121050.3121074}

\bibitem{puschmann2019BubbleAssessingDiversity}
Puschmann, C.: Beyond the bubble: Assessing the diversity of political search results. Digital Journalism  \textbf{7}(6),  824--843 (Jul 2019). \doi{10.1080/21670811.2018.1539626}

\bibitem{putra2018searchx}
Putra, S.R., Moraes, F., Hauff, C.: Searchx: Empowering collaborative search research. In: The 41st International ACM SIGIR Conference on Research \& Development in Information Retrieval. pp. 1265--1268 (2018)

\bibitem{reuver2021StanceDetectionTopicIndependent}
Reuver, M., Verberne, S., Morante, R., Fokkens, A.: Is stance detection topic-independent and cross-topic generalizable? - a reproduction study. In: Proceedings of the 8th Workshop on Argument Mining. pp. 46--56. Association for Computational Linguistics, Punta Cana, Dominican Republic (Nov 2021). \doi{10.18653/v1/2021.argmining-1.5}

\bibitem{Ribeiro2016WhySI}
Ribeiro, M.T., Singh, S., Guestrin, C.: “why should i trust you?”: Explaining the predictions of any classifier. Proceedings of the 22nd ACM SIGKDD International Conference on Knowledge Discovery and Data Mining  (2016)

\bibitem{rieger2021ThisItemMight}
Rieger, A., Draws, T., Tintarev, N., Theune, M.: This {Item} {Might} {Reinforce} {Your} {Opinion}: {Obfuscation} and {Labeling} of {Search} {Results} to {Mitigate} {Confirmation} {Bias}. In: Proceedings of the 32nd {ACM} {Conference} on {Hypertext} and {Social} {Media}. pp. 189--199. {HT} '21, Association for Computing Machinery, New York, NY, USA (2021). \doi{10.1145/3465336.3475101}, \url{https://doi.org/10.1145/3465336.3475101}

\bibitem{roy2022ExploitingStanceHierarchies}
Roy, A., Fafalios, P., Ekbal, A., Zhu, X., Dietze, S.: Exploiting stance hierarchies for cost-sensitive stance detection of web documents. J. Intell. Inf. Syst.  \textbf{58}(1),  1–19 (feb 2022). \doi{10.1007/s10844-021-00642-z}, \url{https://doi.org/10.1007/s10844-021-00642-z}

\bibitem{Sanh2019DistilBERTAD}
Sanh, V., Debut, L., Chaumond, J., Wolf, T.: Distilbert, a distilled version of bert: smaller, faster, cheaper and lighter. ArXiv  \textbf{abs/1910.01108} (2019)

\bibitem{sepulveda2021exploring}
Sep{\'u}lveda-Torres, R., Vicente, M., Saquete, E., Lloret, E., Palomar, M.: Exploring summarization to enhance headline stance detection. In: International Conference on Applications of Natural Language to Information Systems. pp. 243--254. Springer (2021)

\bibitem{silalahi2022named}
Silalahi, S., Ahmad, T., Studiawan, H.: Named entity recognition for drone forensic using bert and distilbert. In: 2022 International Conference on Data Science and Its Applications (ICoDSA). pp. 53--58. IEEE (2022)

\bibitem{srivastava2014dropout}
Srivastava, N., Hinton, G., Krizhevsky, A., Sutskever, I., Salakhutdinov, R.: Dropout: a simple way to prevent neural networks from overfitting. The journal of machine learning research  \textbf{15}(1),  1929--1958 (2014)

\bibitem{staliunaite2020compositional}
Stali{\=u}nait{\.e}, I., Iacobacci, I.: Compositional and lexical semantics in roberta, bert and distilbert: a case study on coqa. arXiv preprint arXiv:2009.08257  (2020)

\bibitem{IG_Sundararajan2017AxiomaticAF}
Sundararajan, M., Taly, A., Yan, Q.: Axiomatic attribution for deep networks. In: International Conference on Machine Learning (2017)

\bibitem{tong2022multimodel}
Tong, J., Wang, Z., Rui, X., et~al.: A multimodel-based deep learning framework for short text multiclass classification with the imbalanced and extremely small data set. Computational Intelligence and Neuroscience  \textbf{2022} (2022)

\bibitem{white2013BeliefsBiasesWeb}
White, R.: Beliefs and biases in web search. In: Proceedings of the 36th International ACM SIGIR Conference on Research and Development in Information Retrieval. pp. 3--12. ACM, Dublin Ireland (Jul 2013). \doi{10.1145/2484028.2484053}

\bibitem{wolf-etal-2020-transformers}
Wolf, T., Debut, L., Sanh, V., Chaumond, J., Delangue, C., Moi, A., Cistac, P., Rault, T., Louf, R., Funtowicz, M., Davison, J., Shleifer, S., von Platen, P., Ma, C., Jernite, Y., Plu, J., Xu, C., Scao, T.L., Gugger, S., Drame, M., Lhoest, Q., Rush, A.M.: Transformers: State-of-the-art natural language processing. In: Proceedings of the 2020 Conference on Empirical Methods in Natural Language Processing: System Demonstrations. pp. 38--45. Association for Computational Linguistics, Online (Oct 2020), \url{https://www.aclweb.org/anthology/2020.emnlp-demos.6}

\bibitem{xu-etal-2018-cross}
Xu, C., Paris, C., Nepal, S., Sparks, R.: Cross-target stance classification with self-attention networks. In: Proceedings of the 56th Annual Meeting of the Association for Computational Linguistics (Volume 2: Short Papers). pp. 778--783. Association for Computational Linguistics, Melbourne, Australia (Jul 2018). \doi{10.18653/v1/P18-2123}, \url{https://aclanthology.org/P18-2123}

\bibitem{yang2017MeasuringFairnessRanked}
Yang, K., Stoyanovich, J.: Measuring fairness in ranked outputs. In: Proceedings of the 29th International Conference on Scientific and Statistical Database Management. pp.~1--6. ACM, Chicago IL USA (Jun 2017). \doi{10.1145/3085504.3085526}

\bibitem{ying2019overview}
Ying, X.: An overview of overfitting and its solutions. In: Journal of physics: Conference series. vol.~1168, p. 022022. IOP Publishing (2019)

\bibitem{Yu2022leg}
Yu, P., Rahimi, R., Allan, J.: Towards explainable search results: A listwise explanation generator. In: Proceedings of the 45th International ACM SIGIR Conference on Research and Development in Information Retrieval. p. 669–680. SIGIR '22, Association for Computing Machinery, New York, NY, USA (2022). \doi{10.1145/3477495.3532067}, \url{https://doi.org/10.1145/3477495.3532067}

\bibitem{zehlike2021FairnessRankingSurvey}
Zehlike, M., Yang, K., Stoyanovich, J.: Fairness in ranking: A survey. arXiv:2103.14000 [cs]  (May 2021), \url{http://arxiv.org/abs/2103.14000}

\end{thebibliography}
\end{document}